\documentclass[aps,prb,twocolumn,superscriptaddress,showpacs]{revtex4-1}
\usepackage{graphicx,color}

\begin{document}


\title{{ NMR study of thermally activated paramagnetism in metallic low-silica X zeolite filled with sodium atoms}}

\author{Mutsuo Igarashi}
\email{e-mail: igarashi@nat.gunma-ct.ac.jp}
\affiliation{Department of Applied Physics, Gunma National College of Technology, Toribamachi 580, Maebashi 371-8530, Gunma, Japan}

\author{Takehito Nakano}
\affiliation{Department of Physics, Graduate School of Science, Osaka University, Toyonaka 560-0043, Osaka, Japan}

\author{Pham Tan Thi}
\affiliation{Department of Physics, Graduate School of Science, Osaka University, Toyonaka 560-0043, Osaka, Japan}

\author{Yasuo Nozue}
\affiliation{Department of Physics, Graduate School of Science, Osaka University, Toyonaka 560-0043, Osaka, Japan}

\author{Atsushi Goto}
\affiliation{National Institute for Materials Science, Sakura 3-13, Tsukuba 305-0003, Ibaraki, Japan}

\author{Kenjiro Hashi}
\affiliation{National Institute for Materials Science, Sakura 3-13, Tsukuba 305-0003, Ibaraki, Japan}

\author{Shinobu Ohki}
\affiliation{National Institute for Materials Science, Sakura 3-13, Tsukuba 305-0003, Ibaraki, Japan}

\author{Tadashi Shimizu}
\affiliation{National Institute for Materials Science, Sakura 3-13, Tsukuba 305-0003, Ibaraki, Japan}

\author{Andra\v{z} Krajnc}
\affiliation{Jo\v{z}ef Stefan Institute, Jamova 39, 1000 Ljubljana, Slovenia}

\author{Peter Jegli\v{c}}
\affiliation{Jo\v{z}ef Stefan Institute, Jamova 39, 1000 Ljubljana, Slovenia}
\affiliation{EN-FIST Centre of Excellence, Dunajska 156, 1000 Ljubljana,
Slovenia}

\author{Denis Ar\v{c}on}
\affiliation{Jo\v{z}ef Stefan Institute, Jamova 39, 1000 Ljubljana, Slovenia}
\affiliation{Faculty of Mathematics and Physics, University of Ljubljana, Jadranska 19, 1000 Ljubljana, Slovenia}

\date{\today}

\begin{abstract}
{ We report a $^{23}$Na and $^{27}$Al nuclear magnetic resonance (NMR) investigation of low-silica~X (LSX) zeolite with chemical formula Na$_{12}$Al$_{12}$Si$_{12}$O$_{48}$ (Na$_{12}$-LSX) loaded with $n$ additional guest sodium atoms.
Na$_n$/Na$_{12}$-LSX exhibits an insulator-to-metal transition around $n=11.6$, which is accompanied by a significant enhancement of bulk magnetic susceptibility. 
Paramagnetic moments are thermally activated in the metallic Na$_{12}$/Na$_{12}$-LSX with an activation energy of around 0.1~eV. 
At the same time, a new shifted component (SC) appears in the $^{23}$Na NMR, whose large and positive NMR shift scales with bulk magnetic susceptibility. 
Its spin-lattice relaxation rate $1/T_1$ is governed by the fluctuations determined by the same activation energy as obtained from the bulk magnetic susceptibility data.
The timescale of these fluctuations is typical for atomic motions, which suggest strong electron-phonon coupling, a hallmark of polaron states. 
The insulator-to-metal transition in Na$_n$/Na$_{12}$-LSX is thus discussed within a polaron model. 
}
\end{abstract}

\pacs{71.20.Dg, 71.30.+h, 75.20.-g, 76.60.-k, 76.60.Cq , 82.75.Vx}

\maketitle
%
%
\section{\label{sec:level1}Introduction}
%
%
Zeolites are nanoporous materials with periodic nanospaces (cages) formed by a covalently bonded framework. 
These cages can accommodate various atoms and molecules, which is widely exploited in catalysis, separations, and ion exchange processes.\cite{Breck_DW_1974} 
{On the other hand, the zeolites with absorbed alkali metal atoms exhibit intriguing magnetic properties ranging from a ferromagnetic,\cite{Nozue_Y_1992} antiferromagnetic,\cite{Srdanov_VI_1998, Damjanovic_L_2000} or even ferrimagnetic orderings.\cite{Nakano_T_2006, Hanh_DT_2010} 
%
%
Moreover, metallic behavior and an insulator-to-metal transition as a function of alkali metal loading have been reported in low-silica X (LSX) zeolites filled with sodium and zeolites Rho filled with rubidium.\cite{Nakano_T_2010, Anderson_PA_2004} 
\par
In these electromagnetically interesting materials, which contain only aluminosilicates and alkali metal atoms, alkali metals tend to form nanoclusters in zeolite cages. 
Some of them display low--temperature magnetic orderings both in the insulating and metallic states.\cite{Nozue_Y_1992, Srdanov_VI_1998, Damjanovic_L_2000, Nakano_T_2006, Hanh_DT_2010, Nozue_Y_1993, Nakano_T_2007, Duan_TC_2007, Duan_TC_2007-2, Duan_TC_2009} 
The appearance of magnetic moments and their ordering has been a great surprise as all the individual components are nonmagnetic. 
The first-principles band-structure calculations within the local-density approximation showed that the alkali-doped zeolites can be described by a tight-binding band of ``superatoms'' -- the states formed at the alkali-metal nano clusters.\cite{Arita_R_2004} 
In this respect zeolites thus behave quite differently from the other alkali metal intercalated compounds, e.g. such as C$_{60}$ or graphite, \cite{Gunnarsson} where alkali atoms play a role of electron donors to the framework and the observed electronic properties arise from the electronic states of the host structure.
\par
In the lattice of alkali-metal nanoclusters with electronic energy band partially occupied, naively a metallic state is expected. 
However, many zeolites show insulating ground state in the wide range of experimentally accessible loading density.
A representative example is zeolite A loaded with potassium which shows ferromagnetic properties \cite{Nozue_Y_1992, Nozue_Y_1993, Nakano_T_2007} and is in insulating state in the full range of loading density.\cite{Nakano_T_1999} 
An important energy scale in such lattice of nanoclusters is the on-site Coulomb repulsion, which in certain cases prevails over the bandwidth and pushes the system to the Mott-insulating state.\cite{Nakano_T_1999, Arita_R_2004} 
On the other hand, the metallic state has been proved in sodium-doped LSX zeolite  at high sodium loading, $n$, after the appearance of the Drude term in the infrared spectrum\cite{Nakano_T_2010} and the dramatic resistivity drop at room temperature for $n\geq 11.6$ as well as by the nearly temperature independent resistivity down to 50~K for $n=12$.\cite{Nozue_Y_2012} 
Another important concurrent energy scale is the electron-phonon coupling. 
The strong electron-phonon coupling of Na atoms, which overcomes the Coulomb repulsion, can stabilize the spin-singlet state. 
This is believed to be the case in the zeolite A loaded with sodium where a nonmagentic insulating state is robust to any loading density.\cite{Kodaira_T_1996} 
%
\par
Various possible arrangements of the zeolite framework that result in the variation of cage potential depths and crystal field distribution lead to the rich variety of electronic states, which depend on the exact local atomic configuration of alkali metal atoms within the zeolite framework.
The confined geometry of alkali-metal nanoclusters imposed by framework cages may also additionally enhance coupling between the electronic and lattice degrees of freedom leading to the formation of polaron states.\cite{Nozue_Y_2012} 
These polarons may be either localized in cages, small polarons, or form extended states over several cages as it is with large polarons. 
This directly opens important questions about the real origin of unquenched spin often found in these materials and the true nature of the metallic state. 
Motived by the above issues we decided to focus on Na-doped LSX zeolite -- a representative system that exhibits an insulator-to-metal transition and the paramagnetism in the metallic state.\cite{Nozue_Y_2012} 
We employed a local probe $^{23}$Na and $^{27}$Al  nuclear magnetic resonance (NMR) technique and found strong indications for the electron-phonon coupling that directly affects the localization and the magnetic moment of Na-cluster states. 
 We show that all NMR data are consistent with the polaron model proposed recently.\cite{Nozue_Y_2012} 
%
%
%
%
\begin{figure}[b]
\includegraphics[width=1.0\linewidth]{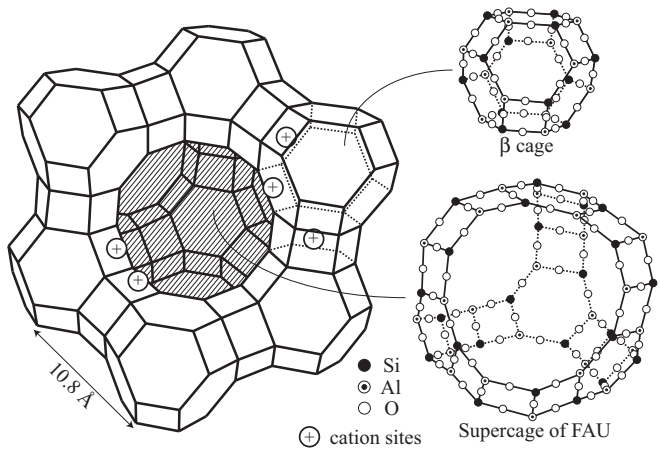}
\caption{\label{fig_crystal} Schematic { structure} of the FAU zeolite { framework}. The small circles with a plus { sign} indicate possible cation sites.}
\end{figure}
%
%
%
%
\section{Experimental}
%
The LSX is an aluminosilicate zeolite with a faujasite (FAU) structure.\cite{Feuerstein_M_1998} 
In the FAU structure, $\beta$ cages with an inner diameter of $\sim 7$~\AA~{ form} a diamond-type structure, as shown in Fig.~\ref{fig_crystal}.
Supercages surrounded by the $\beta$ cages have the inner diameters of $\sim 13$~\AA~and are also arranged in the diamond-type structure. 
The FAU structure has a chemical formula A$_{m}$Al$_{m}$Si$_{24-m}$O$_{48}$ per one supercage, where A represents alkali cations necessary for charge compensation. 
The unit cell has a lattice constant of $\sim 25$~\AA. 
Charge-compensating cations are distributed within the framework as indicated in Fig.~\ref{fig_crystal}.
The LSX zeolite has exactly $m=12$ cations with Si/Al ratio equal to 1. 
For LSX, the so called L\"owenstein's rule\cite{Loewenstein_W_1954} is applicable resulting in alternating arrangement of Si and Al atoms. 
Therefore, the LSX framework is structurally quite ordered compared to other more disordered FAU zeolites with higher Si/Al ratio. 
The absence of structural disorder makes LSX zeolites perfect candidates for the studies of insulator-to-metal transitions as a function of alkali-metal doping. 
\par
The LSX zeolites were loaded with $n$ additional guest sodium atoms per supercage as follows. 
Untreated LSX powder was used as the raw material for sample preparation. 
We exchanged all the cations in the cages with sodium by using an aqueous solution of NaCl. 
The resulting powder has the chemical formula Na$_{12}$Al$_{12}$Si$_{12}$O$_{48}$, or shortly Na$_{12}$-LSX. 
Although the obtained powders were dried in air, the cages still contained a large amount of water, which was removed by heating under dynamic vacuum. 
The Na$_{12}$-LSX powder that has been subjected to this drying process is hereafter referred as Na$_{0}$/Na$_{12}$-LSX. 
Sodium metal was loaded into the system from the vapor state by heating it to 473 K in a quartz tube in a furnace. 
Finally, we obtained Na$_{n}$/Na$_{12}$-LSX powders, where $n$ expresses the number of Na atoms per supercage, as already explained above. 
To compare the insulating and metallic properties of these materials, we prepared samples with $n=10$ and $12$ on the either side of the insulator-to-metal boundary. 
The latter of the two samples corresponds to the maximum possible doping level. 
\par
The temperature dependence of bulk magnetic susceptibility, $\chi$, was measured with a Quantum Design MPMS-XL system. 
Measurements were conducted on cooling from room temperature to 2.0~K in a magnetic field of 5 T. 
\par
For NMR measurements the samples were transferred into the quartz tubes without being exposed to air. 
A small amount of helium gas was added for good thermal contact between the sample and the cryostat.
$^{23}$Na ($I=3/2$) and $^{27}$Al ($I=5/2$) NMR spectra were measured in a magnetic field of 9.4 T in the temperature range between 4 and 300~K.
The reference frequencies of $\nu (^{23}{\rm Na}) = 105.86~$MHz and $\nu (^{27}{\rm Al}) = 104.28~$MHz were determined from NaCl and AlCl$_3$ aqueous solution standards, respectively. 
The same receiver gain was used at all temperatures to follow the temperature variation of the signal intensity. 
Additional experiments were performed in magnetic fields of 6.3, 4.7, and 2.35~T to examine the field dependence of the spin-lattice relaxation rate. 
%
\section{Results}
%
%
\subsection{Magnetic Susceptibility}
%
%
%
%
\begin{figure}[htbp]
\includegraphics[width=0.9\linewidth]{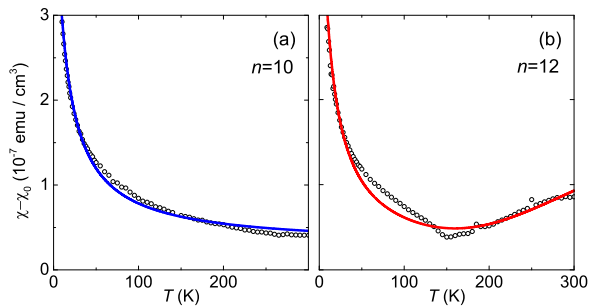}
\caption{
\label{fig_chi_vs_T_merg}
(Color online) Temperature dependence of bulk magnetic susceptibility, $\chi$, of Na$_n$/Na$_{12}$-LSX powder samples with (a) $n=10$ and (b) $n=12$. 
Solid lines represent fits to the experimental data using Eqs.~(\ref{eq_chi(n=10)}) and (\ref{eq_chi(n=12)}). 
The $\chi$ values are corrected by subtracting the diamagnetic contribution $\chi _0 = -1.13 \times 10^{-6}$~emu/cm$^3$ and $-0.75 \times 10^{-6}$~emu/cm$^3$, respectively. 
The major contributions to $\chi _0$'s are from the quartz sample holders. 
}
\end{figure}
Fig.~\ref{fig_chi_vs_T_merg} shows the temperature dependence of the bulk magnetic susceptibility for Na$_n$/Na$_{12}$-LSX samples with $n=10$ and $n=12$. 
Their overall temperature dependences are similar to those reported in a previous study.\cite{Nozue_Y_2012} 
For $n=10$, $\chi$ follows the Curie--Weiss susceptibility, $\chi_{\rm CW}$, for all temperatures, 
\begin{eqnarray}
\label{eq_chi(n=10)}
\chi = \chi_{\rm CW}
=  
\frac
{N_{\rm CW} g^2 \mu_{B}^2 S (S+1)}
{3k_{B} (T - T_{\rm W})}. 
\end{eqnarray}
\noindent
Here $k_{B}$ and $\mu_{B}$ are the Boltzmann constant and the Bohr magneton, respectively. 
Assuming $S=1/2$ and $g=2$ (based on a preliminary electron paramagnetic resonance data), best fit to Eq.~(\ref{eq_chi(n=10)}) results in the paramagnetic spin density $N_{\rm CW} = (8.9\pm 0.2) \times 10^{18} $~cm$^{-3}$ and the small Weiss temperature $T_{\rm W} = -10.7\pm0.4$~K. 
The value of $N_{\rm CW}$ represents only $1.7\pm 0.1 $~\% of the supercage density ($5.12 \times 10^{20}$~cm$^{-3}$). 
Therefore, these paramagnetic spins { probably} arise from a relatively small number of paramagnetic clusters or impurities. 
We conclude that the sample with $n=10$ is intrinsically non-magnetic.
\par
In the sample with $n=12$, a similar Curie-Weiss susceptibility is observed at low temperatures. 
However, above 150~K the temperature dependence of $\chi$ dramatically changes as it suddenly becomes progressively enhanced with increasing temperature. 
This observation suggests the existence of an energy gap leading to the thermally activated behavior, in striking contrast to the $n=10$ sample. 
\par
To examine the thermally activated component, we fit the data using the following { phenomenological model},
\begin{equation}
\label{eq_chi(n=12)}
\chi = \chi_{\rm CW} + \chi_{\rm T},
\end{equation}
where
\begin{equation}
\nonumber
\chi _{\rm T}
=
\frac
{N_{\rm T} g^2 \mu_{B}^2}
{4k_{B} T}
\exp \left(-\frac {\Delta} {k_{B} T} \right).
\end{equation}
Here, $N_{\rm T}$ denotes the density of thermally activated magnetic moments. 
We assume that the number of paramagnetic spins obeys the Arrhenius dependence with an activation energy $\Delta$.
Although the fit is not perfect, the general trend of the temperature dependence of $\chi$ is correctly reproduced (Fig.~\ref{fig_chi_vs_T_merg}).
$N_{\rm T}$ is by more than two orders of magnitude larger than $N_{\rm CW}$, which indicates that every supercage is filled with $S=1/2$ magnetic moment.
This fact implies that $\chi_{\rm T}$ is indeed an intrinsic susceptibility for the metallic $n=12$ sample. 
The metallic state magnetic susceptibility is thus characterized by the thermally activated behavior with an activation energy of around $\Delta \sim 0.1$~eV. 
%
%
\subsection{NMR Spectra}
%
%
%
%
\begin{figure}[tbp]
\includegraphics[width=0.9\linewidth]{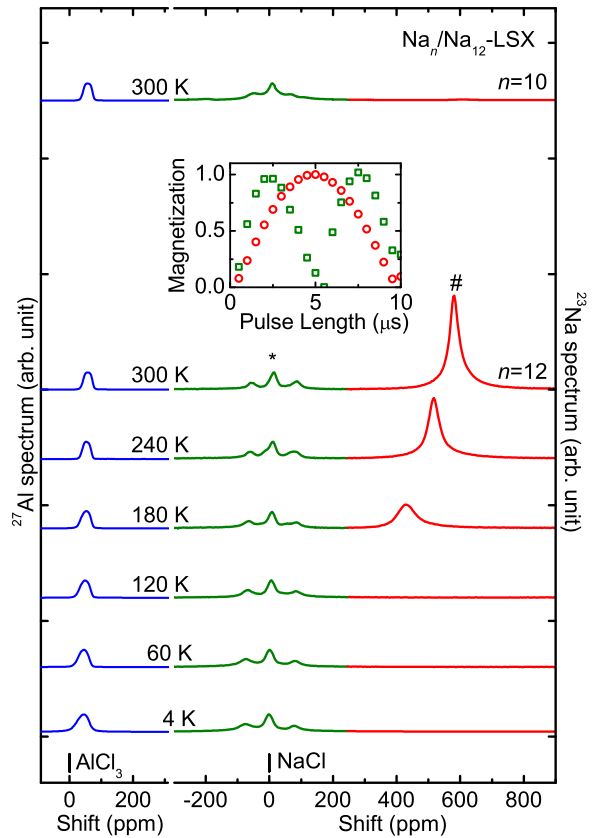}
\caption{
\label{fig_GSpcTvar}
(Color online)
Temperature dependence of $^{27}$Al (left) and $^{23}$Na (right) NMR spectra in Na$_{12}$/Na$_{12}$-LSX. Strongly shifted (SC) and residual (RC) components of the $^{23}$Na NMR spectra are marked with * and \#, respectively. The $^{23}$Na peak heights are at each temperature normalized according to the intensity of the RC component. For comparison room temperature $^{27}$Al and $^{23}$Na  NMR spectra of Na$_{10}$/Na$_{12}$-LSX sample are shown at the top. Please note in this case the complete absence of SC component in the $^{23}$Na spectrum. 
Inset: the normalized nuclear magnetization versus pulse length at 300~K for SC (green circles) and RC (red squares).
}
\end{figure}
%
%
%
%
\begin{figure}[b]
\includegraphics[width=0.8\linewidth]{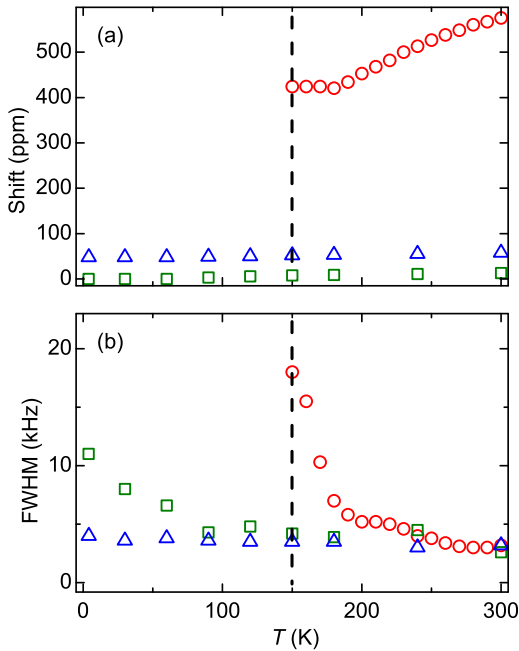}
\caption{
\label{fig_Sft_FWHM_Inten}
(Color online)
{ Temperature dependence of (a) peak shift and (b) the full width at half maximum (FWHM), evaluated for $^{23}$Na SC (red circles), $^{23}$Na RC (green squares), and $^{27}$Al line (blue triangles). 
The SC appears only above 150~K.} 
}
\end{figure}
Representative $^{23}$Na NMR spectra of Na$_{12}$/Na$_{12}$-LSX in a magnetic field of 9.4 T are shown in Fig.~\ref{fig_GSpcTvar} between room temperature and 4~K. 
A spectral component, which is a superposition of three peaks centered close to the $^{23}$Na Larmor frequency, is observed at all temperatures. 
A similar component, which is hereafter denoted as the residual component (RC), was { previously} observed in Na$_{0}$/Na$_{12}$-LSX (Not shown) and now also in Na$_{10}$/Na$_{12}$-LSX (Fig.~\ref{fig_GSpcTvar}). 
No signal was observed at the position corresponding to bulk sodium metal, which has a { temperature-independent} shift of approximately 0.12\%.\cite{Carter_GC_1977}
We thus conclude that there is no residual unreacted metallic sodium present in our samples. 
\par
A new strongly shifted component (SC) is observed only in the metallic Na$_{12}$/Na$_{12}$-LSX sample at approximately 580 ppm at 300~K. 
Interestingly, the SC has the optimal $\pi /2$ pulse length precisely two times longer compared to RC (inset to Fig.~\ref{fig_GSpcTvar}). 
We can thus immediately conclude that the electric field gradient (EFG) at the sodium sites contributing to the SC signal is averaged out to zero,\cite{Man_PP_1988} presumably due to their own fast motion on the NMR time-scale. 
Contrary, the sodium sites contributing to the RC signal experience a nonzero EFG and are thus excited with shorter pulses. 
Another important distinction between the SC and RC is that} the SC has a { significantly shorter} spin-spin relaxation time $T_2$ (68~$\mu$s at 300~K) compared to RC (810~$\mu$s). 
This indicates rapid fluctuations of the local magnetic fields at the sodium nuclei contributing to the SC signal also corroborating with its large shift. 
\par
In addition to the $^{23}$Na NMR signal, the $^{27}$Al spectra of Na$_{12}$/Na$_{12}$-LSX are shown on the left side of Fig.~\ref{fig_GSpcTvar}.
The $^{27}$Al spectra are almost temperature independent and always found close to the corresponding Larmor frequency. 
Only a gradual increase of the linewidth with decreasing temperature is observed. 
Therefore, the electron spin density at the framework Al sites must be negligibly small. 
We finally stress, that the frequency scans of several percents above and below the $^{27}$Al Larmor frequency did not reveal any additional spectral component. 
\par
However, if the above-mentioned SC is actually the $^{27}$Al signal then the shift of this line would be enormous, i.e. approximately 1.57\%. 
Taking then the same hyperfine coupling constant as in  bulk aluminium metal where NMR shift amounts 0.16\% (Ref. \onlinecite{Carter_GC_1977}) the shift of 1.57\% would require 10 times higher density of states (DOS) at the Fermi energy. 
This seems to directly contradict with a relatively low conductivity of the Na$_{12}$/Na$_{12}$-LSX material,\cite{Nozue_Y_2012} which implies low DOS. 
The SC thus cannot be $^{27}$Al resonance and we therefore attribute it to the signal of $^{23}$Na. 
Strong shift of SC resonance then firmly proves that the electronic spin density is on Na-clusters and not on aluminosilicate framework. 
\par
Fig.~\ref{fig_Sft_FWHM_Inten} summarizes the temperature dependences of the peak shift and the full width at half maximum (FWHM) for SC, RC and $^{27}$Al NMR signals, respectively. 
$^{27}$Al and RC NMR signals display a very weak peak-shift and FWHM  temperature dependences and seem not to be sensitive to dramatic changes in $\chi$ around 150 K. 
On the other hand, SC shows much more pronounced temperature dependences: 
FWHM increases in a nearly divergent manner with decreasing temperature, whereas the shift monotonically decreases with decreasing temperature. 
Approaching 150~K, i.e. the temperature where $\chi$ has a minimum, SC line significantly broadens while its shift saturates at 420 ppm.
\par
\begin{figure}[tb]
\includegraphics[width=1\linewidth]{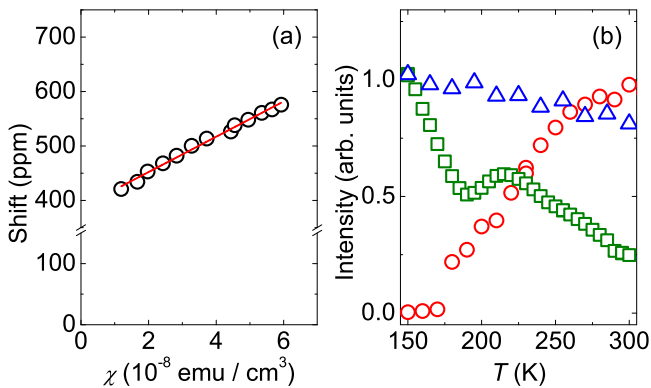}
\caption{
\label{fig_Itn_T_and_fig_G_K-chi}
(Color online) 
(a) 
$^{23}$Na NMR shift of SC versus bulk magnetic susceptibility corrected for the diamagnetic contribution, $\chi-\chi_0$ in Na$_{12}$/Na$_{12}$-LSX. Solid line is a fit to $K=A\chi$ yielding $A=13.7$~kOe/$\mu_B$. 
(b) 
 Temperature dependence of the spectral intensity corrected by the Boltzmann factor for $^{23}$Na SC (red circles), $^{23}$Na RC (green squares), and $^{27}$Al line (blue triangles) in Na$_{12}$/Na$_{12}$-LSX.
}
\end{figure}

The $^{23}$Na NMR spectra thus closely reflect changes in the bulk magnetic susceptibility (Fig.~\ref{fig_chi_vs_T_merg}) above 150 K: the SC is observed for $n=12$ where enhanced and thermally activated $\chi$ has been measured, whereas it is completely absent for $n=10$ where $\chi$ is dominated by the impurity states. 
The large and positive $^{23}$Na NMR shift, $K_{\rm s}$, of SC is thus associated with the hyperfine interaction between $^{23}$Na nuclei and the thermally activated magnetic moments.
To test this possibility we show in Fig.~\ref{fig_Itn_T_and_fig_G_K-chi} the so-called Clogston-Jaccarino plot\cite{Slichter} where linear dependence $K_{\rm s}=A\chi$ speaks firmly in favor of the proposed hyperfine coupling. 
If magnetic moments are homogeneously distributed among supercages, then the hyperfine coupling $A$ equals 13.7~kOe/$\mu_B$. 
\par
The SC resonance completely disappears below 150~K in present experiments as demonstrated by its Boltzmann-corrected intensity (Fig.~\ref{fig_Itn_T_and_fig_G_K-chi}b). 
This surprising behavior of SC is reversible without any observable temperature hysteresis. 
In the same temperature range the intensity of RC increases with decreasing temperature, whereas for $^{27}$Al it remains almost constant, as expected. 
Although the loss in SC is not completely compensated by the corresponding increase in RC, a clear correlation between the intensity of SC line and the thermally activated magnetic moments can be firmly established. 
%
%
%
%
%
%
%
\subsection{Spin--Lattice Relaxation}
%
%
%
%
\begin{figure}[htb]
\includegraphics[width=1\linewidth]{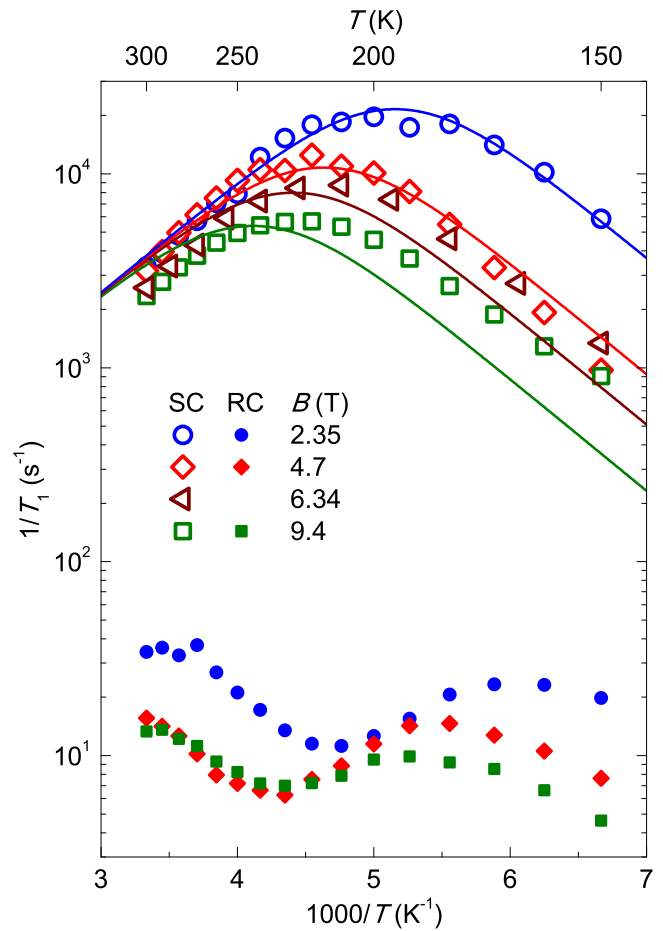}
\caption{
\label{fig_lnT1_T}
(Color online) { Temperature dependence of the $^{23}$Na $T_1^{-1}$ for SC { (open symbols)} and RC { (solid symbols)} measured in different magnetic fields{ : 2.35 T (circles), 4.7 T (diamonds), 6.34 T (triangles), and 9.4 T (squares)}. Solid lines represent a global fit with the BPP model, { Eqs.~(\ref{BPP}) and (\ref{expEa})}.}
}
\end{figure}
We now consider the $^{23}$Na spin-lattice relaxation mechanism of the SC. 
The $^{23}$Na magnetization curves exhibit a rapid but still clearly observable recovery and are well described by a single exponential function (not shown). 
The extracted spin-lattice relaxation rates, $T_1^{-1}$, of SC measured in different magnetic fields are summarized in Fig.~\ref{fig_lnT1_T}. 
Strongly non-monotonic temperature dependence of $T_1^{-1}$ is incompatible with the Korringa relation 
\begin{equation}
\label{Korringa}
T_1TK_{\rm s}^2 = \frac{\hbar}{4\pi k_B}\frac{\gamma_e^2}{\gamma_I^2},
\end{equation}
expected for normal metals. 
Here, $\gamma_e$ and $\gamma_I$ represent the electron and nuclear gyromagnetic ratios, respectively. 
In addition, using the experimental values measured at 300~K ($K_{\rm s}=580$~ppm and $T_1=0.40$~ms), we calculate $T_1 T K_{\rm s}^2 \approx 4 \times 10^{-8}$~sK, which is by a factor of 100 less than the value of $3.8 \times 10^{-6}$~sK anticipated from the Korringa relation. 
This simple estimate strongly suggests that in metallic Na$_{12}$/Na$_{12}$-LSX spin-lattice relaxation via conducting electrons is not the dominant relaxation channel and that other relaxation mechanism must be considered too. 
\par
In fact, the observed temperature dependence resembles the Bloembergen-Purcell-Pound (BPP)-type mechanism\cite{BPP_1948} with 
\begin{equation}
\label{BPP} 
T_{1}^{-1}=C \frac{\tau_c}{1+\omega^2\tau_c^2}, 
\end{equation}
\noindent
where $\tau_c$ is the correlation time for the local field fluctuations at the nucleus, $\omega$ is the Larmor angular frequency and $C$ is a measure of the fluctuating local fields magnitude. 
We employ a simplified version of BPP relaxation for $^{23}$Na nuclei, since the exact nature of interaction with the source of fluctuations is not known {\em a priori}. 
Assuming that fluctuations exhibit the Arrhenius behavior,\cite{Slak_J1984} 
we set 
\begin{equation}
\label{expEa} 
\tau_c=\tau_0\exp\left( \frac{E_a}{k_B T} \right), 
\end{equation}
\noindent
where $\tau_0$ is a constant, $E_a$ is the activation energy for the field fluctuations, and $k_B$ is the Boltzmann constant. 
The solid lines in Fig.~\ref{fig_lnT1_T} represent a global fit to the experimental data using Eqs.~(\ref{BPP}) and (\ref{expEa}). 
We obtain $E_a=(0.11 \pm 0.01)$~eV, $\tau_0=(6.3 \pm 1.5)\times 10^{-12}$~s and $C=(7.2 \pm 0.2) \times 10^{12}$~s$^{-2}$. 
Here we stress that different sets of experimental data differ only in the value of $\omega$ determined by the external magnetic field. 
Good agreement between the BPP model and experimental data implies that the relaxation is governed by a thermally activated fluctuations of the local (magnetic or crystal) field at the sodium sites contributing to the SC signal. 
Similar explanation for $T_1^{-1}$ has been given in the case of sodalite,\cite{Heinmaa_I_2000} although the relaxation rates were there several orders of magnitude lower compared to SC in Na$_{12}$/Na$_{12}$-LSX.
\par
On the other hand, the spin-lattice relaxation rate $T_1^{-1}$ of $^{23}$Na RC is much smaller compared to the SC as shown in Fig.~\ref{fig_lnT1_T}. 
Due to observed distribution of relaxation times $T_1$ and rather complicated temperature dependence of $T_1^{-1}$, the analysis of RC requires additional experimental work in the future. 
However, it is clear that the thermally activated fluctuations of the local fields are felt much stronger by the SC than the RC components. 
%
%
\subsection{Discussion}
%
%
In order to understand our experimental results, we first summarize the main findings for the metallic Na$_{12}$/Na$_{12}$-LSX sample: 
(i) The bulk magnetic susceptibility shows that paramagnetic moments are thermally activated with an activation energy of around 0.1~eV; 
(ii) there is no paramagnetic spin density on the framework judging from the absence of $^{27}$Al NMR shift; 
(iii) a new very intense $^{23}$Na shifted component appears in the metallic phase whose electric field gradient is averaged out to zero presumably due to the sodium fast motion; 
(iv) the  large and positive $^{23}$Na  NMR shift for this component scales with bulk magnetic susceptibility; 
(v) the spin-lattice relaxation of $^{23}$Na SC is determined by the local field fluctuations, which have an activation energy again around 0.1~eV; 
and (vi) the timescale of these fluctuations is not typical for electron dynamics (usually in the femtosecond range) but for atomic motion.
\par
Matching of the activation energies that determine the static magnetic properties ($\chi$ and $K$) and dynamic properties ($T_1$) may not be just the coincidence. 
The same paramagnetic moments not only produce time-averaged static local fields but are also responsible for the fluctuations of local fields at the $^{23}$Na sites contributing to the SC signal. 
However, the timescale and the temperature dependence of these fluctuations are more characteristic for large-amplitude atomic motions. 
Thermally activated paramagnetic moments in the metallic Na$_{12}$/Na$_{12}$-LSX are thus strongly correlated with sodium motions or in other words, the electron-phonon coupling must be important in the investigated system. 
\par

This brings us to the polaron model,\cite{Toyozawa_Y_1961} which has been proposed recently,\cite{Nozue_Y_2012} to explain the insulator-to-metal transition in the sodium-doped Na$_{12}$-LSX zeolite. 
In this picture, the nonmagnetic insulating state is explained by self-trapped small polarons. 
In fact, these small polarons form bipolarons, which are in the spin-singlet ($S=0$) state. 
The position of RC resonance  at the $^{23}$Na Larmor frequency in the insulating Na$_{10}$/Na$_{12}$-LSX is thus fully consistent with the formation of such small bipolarons. 
\par
The metallic phase transition is explained by a formation of large polarons and an adiabatic potential was proposed\cite{Nozue_Y_2012} in order to account for the thermally activated paramagnetism observed in bulk magnetic susceptibility.  
However, when lattice deformations that follow charges become large, the electrons become localized by forming small polaron states with non-zero magnetic moment. 
The temperature dependences of SC shift and its intensity suggest that the small polarons are at high temperatures excited from the large polarons with an activation energy around 0.1~eV. 
In principle small polarons are generated and annihilated all the time. 
This could thus account for the local-field fluctuations probed by $^{23}$Na $T_1^{-1}$ on the SC. 
Since a formation of small polarons involves strong lattice deformation, the timescale for this fluctuation could easily fall into the nanosecond regime typical for atomic motions. 
In addition, polaron model could also explain the fact that the $^{23}$Na sites contributing to the SC signal have zero EFG due to their own fast motion on the timescale of NMR experiment. 
Although large shift of SC peak could in principle account for large DOS appearing close to Mott-insulator instability, strong electron-phonon coupling proved by our NMR experiments favors polaron explanation for the insulator-to-metal transition in Na$_{n}$/Na$_{12}$-LSX family. 
In the future it will be interesting to see in Na$_{12}$/Na$_{12}$-LSX sample whether the SC reappears at low temperatures as a new NMR line with nonzero EFG thus probing small polaron states. 
Further NMR studies, including the temperature-dependent nutation experiment, are necessary to resolve this issue. 
Also other experimental techniques, for instance dielectric spectroscopy,\cite{Levstik_APL_2002} should be employed to give an additional support for polaron model. 
%
%
%
%
\section{Conclusion}
%
%
The insulating Na$_{10}$/Na$_{12}$-LSX and the metallic Na$_{12}$/Na$_{12}$-LSX samples were studied by a combined temperature and field dependent $^{23}$Na and $^{27}$Al NMR. The absence of any $^{27}$Al hyperfine shift rules out aluminosilicate framework as the electronically active component. 
The metallic and (para)magnetic behavior is rather associated with the sodium atoms clustering in cages as evidenced by the appearance of a strongly shifted $^{23}$Na SC resonance. 
This resonance exhibits a thermally activated behavior with an activation energy of about 0.1 eV that correlates with the bulk magnetic susceptibility and suggests strong electron-phonon coupling. 
Its temperature dependence can be well explained by a polaron model proposed recently,\cite{Nozue_Y_2012} where thermally activated behavior is associated with the creation/annihilation of localized small polarons from the bath of large polarons. 

%
%
\section*{Acknowledgments}
%
%
We thank Dr. T. Kodaira and Dr. T. Ikeda for providing high-quality zeolite crystals. 
We also thank Mr. S. Tamiya for assistance with the chemical analysis. 
M.~I. expresses special thanks to Prof. K. Tanigaki, Prof. M. Ito and Prof. Y. Maniwa for helpful discussions. 
He is also grateful to Dr. N. Taira for handling the zeolites. 
P.~J. acknowledges M. Klanj\v{s}ek and T. Apih for their advice. 
This study was partially supported by Grants-in-Aid for Scientific Research [No. 15540353(C)], [No. 242440590(A)], and [No. 19051009(Priority Areas)] from the Ministry of Education, Culture, Sports, Science and Technology (MEXT), Japan.
D.A. acknowledges the financial support by the European Union FP7-NMP-2011-EU-Japan project LEMSUPER under contract no. 283214 and the Researcher Exchange Program between Slovenia and Japan supported by Slovenian research agency. 

\end{document}